\documentclass[a4paper,11pt]{article}
\usepackage{jheppub} 

\usepackage{graphicx}
\usepackage{dcolumn}
\usepackage{bm}
\usepackage{color}   

\usepackage{algpseudocode}
\usepackage{algorithm}
\usepackage{physics}

\usepackage{amsmath} 						
\usepackage{amssymb} 						
\usepackage{mathtools}			
\usepackage{mathrsfs}                       
\usepackage{amsthm}
\usepackage{amssymb}
\usepackage{hyperref}
\hypersetup{
    colorlinks = true,
    linkcolor = blue,
    allcolors=blue
    }

\newtheorem{theorem}{Theorem}

\newtheorem{lemma}{Lemma}
\newtheorem{definition}{Definition}

\usepackage[noabbrev, capitalise,  nameinlink]{cleveref}
\crefname{equation}{Eq.}{Eqs.}
\crefname{theorem}{Theo.}{Theorems.}
\crefname{corollary}{Cor.}{Cor.}
\crefname{lemma}{Lemma}{Lemma}
\crefname{definition}{Def.}{Def.}
\crefname{proposition}{Prop.}{Prop.}
\crefname{discussion}{Disc.}{Disc.}
\crefname{example}{Ex.}{Ex.}
\crefname{proof2}{Proof}{Proof}

\title{Measurable Krylov Spaces and Eigenenergy Count in Quantum State Dynamics}

\author{Saud \v{C}indrak$^a$}
\emailAdd{saud.cindrak@tu-ilmenau.de}

\author{Adrian Paschke$^{b,c}$}
\emailAdd{paschke@inf.fu-berlin.de}

\author{Lina Jaurigue$^a$}
\emailAdd{lina.jaurigue@tu-ilmenau.de}

\author{Kathy L\"udge$^a$}
\emailAdd{kathy.luedge@tu-ilmenau.de}
\affiliation{$^a$Institute of Physics, Technische Universit\"at Ilmenau, Germany}
\affiliation{$^b$Institute of Informatics, Freie Universit\"at Berlin, Germany}
\affiliation{$^c$ Fraunhofer Institute for open communication systems, Germany}

\date{\today}

\abstract{
In this work, we propose a quantum-mechanically measurable basis for the computation of spread complexity. Current literature focuses on computing different powers of the Hamiltonian to construct a basis for the Krylov state space and the computation of the spread complexity. We show, through a series of proofs, that time-evolved states with different evolution times can be used to construct an equivalent space to the Krylov state space used in the computation of the spread complexity. Afterwards, we introduce the effective dimension, which is upper-bounded by the number of pairwise distinct eigenvalues of the Hamiltonian. The computation of the spread complexity requires knowledge of the Hamiltonian and a classical computation of the different powers of the Hamiltonian. The computation of large powers of the Hamiltonian becomes increasingly difficult for large systems. The first part of our work addresses these issues by defining an equivalent space, where the original basis consists of quantum-mechanically measurable states. We demonstrate that a set of different time-evolved states can be used to construct a basis. We subsequently verify the results through numerical analysis, demonstrating that every time-evolved state can be reconstructed using the defined vector space. Based on this new space, we define an upper-bounded effective dimension and analyze its influence on finite-dimensional systems. We further show that the Krylov space dimension is equal to the number of pairwise distinct eigenvalues of the Hamiltonian, enabling a method to determine the number of eigenenergies the system has experimentally. Lastly, we compute the spread complexities of both basis representations and observe almost identical behavior, thus enabling the computation of spread complexities through measurements.
}

\begin{document}

\maketitle

\flushbottom

\section{Introduction}
\label{Introduction}

Our understanding of quantum systems primarily arises from the Schrödinger equation for pure states and the von-Neumann equation for mixed states. The comprehension of how quantum states evolve over time is crucial for various technologies. Therefore, recent years have seen increased attention towards understanding the space in which quantum state evolution occurs.
In this work, we aim to define a measurable space for quantum time evolution under the Schrödinger equation. Our research explores operator complexity, as initially introduced in \cite{PAR19}, and spread complexity, first explored in \cite{BAL22}. A unified perspective on state and operator complexities was later presented in \cite{ALI23}.
Research in the understanding of Krylov complexity has been done in regards to finite and infinite dimensional quantum systems, field theories, quantum chaos, open quantum systems, entanglement, and more (see \cite{AFR23, ANE24, BAE22, BAL22, BAR19, BHA22a, BHA22b, BHA23, BHA24a, CAM23, CAO21, CAP21, CHA23, DYM20, DYM21, FAN22, GUO22A, HAS23, HE22, HEV22, IIZ23, JIA21, KIM22, LI24, LIU23b, MAG20, MUC22, NIZ23, PAT22, RAB22, RAB22a, VAS24, CAP24}). Spread complexity has been analyzed concerning Nielsen complexity, describing topological phases, chaotic dynamics, finite and infinite dimensional quantum systems, and so forth \cite{BAL22,BAL22a, BAL23,CRA23,AGU24,CAP22,CAP22a,ERD23,GIL23,NAN24,PAL23,BHA22,CHA23,GAU23, NIE06a, BHA24, BAL23a, LIN22, RAB23, HUH24}. 

The computation of the spread complexity measures requires classical simulation of the system, which becomes infeasible for larger systems. In this work, we introduce a space that is equivalent to the Krylov space discussed in \cite{BAL22}. We further show that the basis of our space can be computed through quantum mechanical measurements and that the dimension is equal to the number of pairwise distinct eigenvalues of the Hamiltonian. We compute the spread complexity for the introduced basis and compare the result to the spread complexity which is computed using different powers of the Hamiltonian \cite{BAL22}. We show that the spread complexities exhibit almost identical behavior for a simple Ising model. Another interesting aspect of our basis is that the exact Hamiltonian does not need to be known, but just some time-evolved states. This is especially of interest in quantum machine learning, where the quantum systems consist of multiple applications of different Hamiltonians. Additionally, quantum computers attain sizes beyond the computational capabilities of classical machines limiting the usage of krylov complexity for this field.

Quantum computation tries to make use of the exponential scaling in space dimension for computing resources \cite{NIE11}. The most prominent quantum algorithm is Shor's algorithm, which can perform prime factorization more efficiently than a classical computer \cite{SHO94}. The simulation of quantum systems on a quantum computer, which are classically difficult to simulate, appears increasingly plausible with the continuous growth in qubit sizes from 127 qubits in 2021 to 1121 qubits in 2023 on IBM's superconducting devices. Another important aspect of classical computation is machine learning, and therefore, the question arises: how can machine learning algorithms be implemented using quantum systems?

One approach to quantum machine learning is variational quantum machine learning, where a quantum network is trained on a quantum computer. The second scheme is quantum reservoir computing, which tries to utilize the large Hilbert space as a computing resource for time series forecasting. This scheme is especially challenging because of the collapse of the state after measurement. Some solutions to this problem have been proposed in \cite{CIN24, MUJ23}. Lastly, quantum extreme learning machines use quantum systems in a similar way to quantum reservoir computing. The main difference is that the state is reset for each input in quantum extreme learning, whereas it continuously evolves in quantum reservoir computing. Several experiments in quantum machine learning have been conducted on quantum hardware, as detailed in \cite{LI15, SAG21}.

The comparative analysis of machine learning networks remains an active field of research, and the question of expressivity in quantum machine learning (QML) remains unresolved. The central inquiry revolves around the methodology for comparing two Hamiltonians or two quantum machine learning networks in terms of the dimension of the space that comprehensively spans all future states. We propose the effective Krylov dimension as an expressivity measure.

\cref{sec:fund_SC} of this work will introduce the spread complexity, which was proposed in \cite{BAL22}. The development of a measurable Krylov space will be discussed in \cref{sec:meas_KS}. We will prove our main theorems and construct a Krylov space that is computable through quantum mechanical measurements (\cref{theorem:2time_lim} and \cref{theorem:2_2_main_proof}). Afterwards, we will show that the dimension of this space is equivalent to the number of pairwise distinct eigenvalues of the Hamiltonian in \cref{theorem:3Eigenvalue_EQ}. The results will be first verified by the reconstruction capability of the defined space through numerical analysis, see \cref{sec:num_ver}. Here, we will define and discuss the effective Krylov dimension, which is upper-bounded by the number of pairwise distinct eigenenergies of the Hamiltonian. In \cref{sec5:msc}, we will compute the spread complexity using the classical approach and using our introduced basis, where we observe almost identical behavior.

\section{Fundamentals of spread complexity}
\label{sec:fund_SC}
We define the time evolution of a quantum system (\cref{def:1timeEvol}) and the Krylov space (\cref{def:2Krylov}). The observation that the time-evolution operator is a map onto a Krylov space was discussed and \cref{theorem:1EvolKrylov} was proven in \cite{PAR19, BAL22}. The authors further defined a complexity measure based on this discovered property for further analysis of the time evolution in quantum systems.\\
 The Schrödinger equation for the time-independent Hamiltonian $H$ with initial condition $\ket{\Psi(0)}$  is given by:
\begin{align}
    \partial_{t} \ket{\Psi(t)} = -iH\ket{\Psi(t)} \nonumber \\
    \ket{\Psi(0)} := \ket{\Psi_0}
\end{align}
The solution to this equation is
\begin{align}
    \ket{\Psi(t)} = e^{-iHt}\ket{\Psi_0}
\end{align}
and the density matrix $\rho \in \mathbb{C}^{N \times N}$ is constructed by
\begin{align}
    \rho = \ket{\Psi} \bra{\Psi}.
\end{align}
A measurement in quantum mechanics is the application of a Hermitian matrix $O$, also called an observable, with the eigenvalue equation $O\ket{o_i} = o_i\ket{o_i}$ onto the system. Only the eigenvalues of $O$ can be measured, and which eigenvalue is measured is given by the probability
\begin{align}
    P(m=o_i) = |\bra{o_i}\ket{\Psi}|^2.
\end{align}
The expectation value of the observable $O$ in regard to the state $\ket{\Psi}$ is given by
\begin{align}
    \ev{O} = \bra{\Psi} O \ket{\Psi} = \mathrm{Tr}(\rho O).
\end{align}
This work will construct a basis $\ket{w_1}, \ket{w_2}, .. \ket{w_m}$ for some Hamiltonian $H$ and starting state $\ket{\Psi_0}$, such that
\begin{align}
    \ket{\Psi(t)} \in \mathrm{Span}(\ket{w_0}, \ket{w_1}, .., \ket{w_m})
\end{align}
holds for all $t$. 
The established basis should also be attainable through quantum mechanical measurements, particularly for systems where the precise Hamiltonian is unknown or when the system's size surpasses classical simulation capabilities. Quantum mechanics commonly employs the representation of complex vectors within the Hilbert space, denoted as ${\mathcal{H}=\mathbb{C}^N}$.
\begin{definition}[Time Evolution]
\label{def:1timeEvol}
    Let $t\in \mathbb{R}_{\geq 0}$, $\ket{\Psi_0}, \ket{\Psi(t)} \in \mathbb{C}^N$ and $H\in \mathbb{C}^{N\times N}$ be an Hermitian matrix, meaning $H=H^\dagger$. The time evolution $F:\mathbb{R} \times \mathbb{C}^N \rightarrow \mathbb{C}^N$ with initial condition $\ket{\Psi(t=0)}=\ket{\Psi_0}$ is defined by
    \begin{align}
        \ket{\Psi(t)}=F(\ket{\Psi_0})=e^{-iHt}\ket{\Psi_0}.
    \end{align}    
\end{definition}    
\cref{def:2Krylov} gives the general definition of a Krylov space and how to construct this space. 
\begin{definition}[Krylov Space]
\label{def:2Krylov}
Let $f:\mathbb{C}^N \rightarrow \mathbb{C}^N$ be a linear function, i.e., for all $\mathbf{v}, \mathbf{w} \in \mathbb{C}^N$ and $\alpha \in \mathbb{C}$, the following holds:
\begin{align}
    f(\mathbf{v} + \mathbf{w}) &= f(\mathbf{v}) + f(\mathbf{w}), \quad f(\alpha \mathbf{w}) = \alpha f(\mathbf{w})
\end{align}
Let $\mathbf{v}_0 \in \mathbb{C}^N \setminus \{\mathbf{0}\}$ be a vector. Then there exists a uniquely defined smallest $m \leq N$ such that the vectors
\begin{align}
    \mathbf{v}_0, f(\mathbf{v}_0), f^2(\mathbf{v}_0), \ldots, f^{m-1}(\mathbf{v}_0) \nonumber
\end{align}
are linearly independent, and the vectors
\begin{align}
    \mathbf{v}_0, f(\mathbf{v}_0), f^2(\mathbf{v}_0), \ldots, f^{m}(\mathbf{v}_0)
\end{align}
are linearly dependent. 
The number $m$ is called the grade of the vector $\mathbf{\mathbf{v}_0}$ with respect to $f$. The letter $m$ will always be used to denote a grade.
The $j$-th Krylov space of $f$ with respect to $\mathbf{v}_0$ is defined as
\begin{align}
    \mathrm{K}_j(f, \mathbf{v}_0) = \mathrm{Span}\{\mathbf{v}_0, f(\mathbf{v}_0), \ldots, f^{j-1}(\mathbf{v}_0)\} \subseteq \mathbb{C}^N.
    \label{eq:KrylovSpace}
\end{align} 
Note that $\mathrm{K}_{M}(f, \mathbf{\mathbf{v}_0}) = \mathrm{K}_m(f, \mathbf{\mathbf{v}_0})$ holds for all $M \geq m$. If the chosen initial state $\mathbf{v}_0$ and function $f$ are clear, the corresponding Krylov space with grade $m$ will be denoted as $\mathrm{K}_m$.
\end{definition}
\cref{theorem:1EvolKrylov} will give the proof that the time-evolution, as defined in \cref{def:1timeEvol}, maps the initial state $\ket{\Psi_0}$ onto a Krylov space, as discussed in \cite{PAR19,BAL22}.
\begin{theorem}
\label{theorem:1EvolKrylov}
    [Time Evolution as a Map onto a Krylov Space]
    Assume the system from \cref{def:1timeEvol}. Let $t\in \mathbb{R}$ and let  $\ket{\Psi(t)}=\exp(-iHt)\ket{\Psi_0}$ be the time-evolved state by $t$, where $\ket{\Psi_0} \neq 0$. For any $t\in \mathbb{R}$  
    $$\ket{\Psi(t)} \in \mathrm{K}_m(-iH, \ket{\Psi_0})=\mathrm{Span}\{f^0(\ket{\Psi_0}), \ldots, f^{m-1}(\ket{\Psi_0})\}$$
    holds true, where $m \in \mathbb{N}$ is the grade of the starting state $\ket{\Psi_0}$ in regards to the linear function $f:=-iH$\cite{PAR19,BAL22}.
\begin{proof}
        We know that $\ket{\Psi(t)} = \exp(-iHt)\ket{\Psi_0}$ is given. With the introduction of the linear function $f:\mathbb{C}^N\rightarrow \mathbb{C}^N$ as $f(\ket{\Psi_0}) := -iH\ket{\Psi_0}$, the Taylor expansion of the time evolution can be rewritten as
        \begin{align}
            \ket{\Psi(t)} &= F(\ket{\Psi_0}) = e^{-iHt}\ket{\Psi_0} \nonumber \\ 
            &=\sum_{k=0}^\infty (-iH)^k\frac{t^k}{k!}\ket{\Psi_0} = \sum_{k=0}^\infty f^k(\ket{\Psi_0})\frac{t^k}{k!}.
        \end{align}
        $\ket{\Psi(t)}$ is a superposition of vectors $f^k(\ket{\Psi_0})$ with amplitudes $\frac{t^k}{k!}$, such that 
        \begin{align}
            \ket{\Psi(t)} \in \mathrm{Span}\{f^0(\ket{\Psi_0}), f^1(\ket{\Psi_0})t, f^2(\ket{\Psi_0})\frac{t^2}{2!}, \ldots\}
        \end{align}
        holds. In the trivial case with $t=0$, $\ket{\Psi(t)} = \ket{\Psi_0}$ follows. In the case of $t>0$, the values $\frac{t^k}{k!} > 0$, such that the span can be represented as
        \begin{align}
            \mathrm{Span}(f, \ket{\Psi_0}) &= \mathrm{Span}\{f^0(\ket{\Psi_0}), f^1(\ket{\Psi_0}), f^2(\ket{\Psi_0}), \ldots\}.
        \end{align}
        By \cref{def:2Krylov} and the linear property of $f$, there exists an $m\leq N$, such that
        \begin{align}
            &\ket{\Psi_0}, f(\ket{\Psi_0}), f^2(\ket{\Psi_0}), \ldots, f^{m-1}(\ket{\Psi_0}) 
        \end{align}
        are linearly independent, and such that
        \begin{align}
            &\ket{\Psi_0}, f(\ket{\Psi_0}), f^2(\ket{\Psi_0}), \ldots, f^{m}(\ket{\Psi_0}) 
        \end{align}
        are linearly dependent. This implies that
        \begin{align}
            \mathrm{Span}(f, \ket{\Psi_0}) &= \mathrm{Span}\{f^0(\ket{\Psi_0}), f^1(\ket{\Psi_0})t, f^2(\ket{\Psi_0})\frac{t^2}{2!}, \ldots\} \nonumber \\
            &= \mathrm{Span}\{f^0(\ket{\Psi_0}), f^1(\ket{\Psi_0}), f^2(\ket{\Psi_0}), \ldots\} \nonumber \\
            &= \mathrm{Span}\{f^0(\ket{\Psi_0}), f^1(\ket{\Psi_0}), \ldots, f^{m-1}(\ket{\Psi_0})\} \nonumber \\
            &= \mathrm{K}_m(-iH, \ket{\Psi_0}),
        \end{align}
        and therefore $\ket{\Psi(t)} \in \mathrm{K}_m(-iH, \ket{\Psi_0})$, where $m$ is the grade of the vector $\ket{\Psi_0}$ concerning $f:=-iH$.
    \end{proof}

\end{theorem}
In current analysis of quantum complexity the above defined spaces $\mathrm{K_m}$ are used to define a complexity measure \cite{BAL22}. First, the Lanczos algorithm is performed on this space to construct the space 
$\mathrm{K_m}=\mathrm{Span}(\ket{k_0}, \ket{k_2},..,\ket{k_{m-1}})$.
The state representation of any time evolved state $\ket{\Psi(t)}$ is given by
\begin{align}
    \ket{\Psi(t)} = \sum_{n=0}^{m-1}\alpha_n(t)\ket{k_n}.
\end{align}
The spread complexity $C_H$ is am measure of the spread of the state over the basis consisting of different powers of the Hamiltonian and is defined as 
\begin{align}
    C_H(t) = \sum_{n=0}^{m-1} (n+1)\abs{\alpha_{n}(t)}^2.
\end{align}

\section{Measurable Krylov spaces}
\label{sec:meas_KS}
 In our work, we propose a different basis for the Krylov space, which we give \cref{theorem:2time_lim} and \cref{theorem:2_2_main_proof}. We show that instead of constructing the Krylov basis as in \cref{theorem:1EvolKrylov}, the space can be constructed from different time-evolved states of the system. \cref{lemma:1globalPhase} shows that the constructed space is invariant under a global phase, thereby becoming quantum mechanically measurable. With \cref{lemma:equidistantTimes}, we discuss how different evolution times can be chosen to construct the space discussed in \cref{theorem:2time_lim} and \cref{theorem:2_2_main_proof}. To verify the mathematical results, we construct another basis, which implies that the number of pairwise distinct eigenvalues should coincide with the dimension of the Krylov space  (see \cref{theorem:3Eigenvalue_EQ}).
\begin{definition}[Samplings of Krylov Space]
\label{def:3G_Samplings}
Assume $\ket{\Psi_0} \in \mathbb{C}^N$, $\ket{\Psi_0} \neq 0$, and $H \in \mathbb{C}^{N \times N}$, and consider the time evolution for $T_G > 0$ such that 
\begin{align}
    \ket{\Psi(t)} = e^{-iHT}\ket{\Psi_0}
\end{align}
holds. Let $\mathrm{K}_m(-iH, \ket{\Psi_0})$ be the Krylov space of $f:=-iH$ with respect to $\ket{\Psi_0}$, where $m$ is the grade of $\ket{\Psi_0}$ concerning $f:=-iH$. With discrete times $0 = t_0 < t_1 < \ldots < t_{M-1} = T_G$, the vectors
\begin{align}
    \ket{g_i} = \ket{\Psi(t_i)} = e^{-iHt_i}\ket{\Psi_0}\mathrm{,~for~} i = 1, \ldots, M,
\end{align}
are introduced, addressed as the sampling vectors of the time evolution.
\end{definition}
\begin{theorem}
\label{theorem:2time_lim}
    Let $0 = t_0 < t_1 < \ldots < t_{n-1} < T_P$ be some times. Next, the following vectors are introduced:
    \begin{align}
        {h_i^{(n)}}(\ket{\Psi_0}) &:= \sum_{j=0}^{n-1} {f^j(\ket{\Psi_0})}\frac{t_i^j}{j!} \\
        \mathrm{ with~ } f^j(\ket{\Psi_0}) &:= (-iH)^j\ket{\Psi_0} \nonumber.
    \end{align}
    The span of the vectors ${h_i^{(n)}}(\ket{\Psi_0})$ is given by 
    \begin{align}
        \mathrm{H}_n^n &= \mathrm{Span}({h_0^{(n)}}(\ket{\Psi_0}), {h_1^{(n)}}(\ket{\Psi_0}), \ldots, {h_{n-1}^{(n)}}(\ket{\Psi_0})).
    \end{align}
    Then for all $n\in \mathbb{N}$, it holds thatf
    \begin{align}
        \mathrm{H}_n^n = \mathrm{K}_n.
    \end{align}
    \begin{proof}
        Assume $n\in \mathbb{N}$. Then the vectors ${h_i^{(n)}}(\ket{\Psi_0})$ can be represented with the vectors ${f^j(\ket{\Psi_0})}$ as
        \begin{align}
            &{h_i^{(n)}}(\ket{\Psi_0}) = \sum_{j=0}^{n-1}{f^j(\ket{\Psi_0})}\frac{t_i^j}{j!} \nonumber \\
            &= ({f^0(\ket{\Psi_0})},{f^1(\ket{\Psi_0})}, \ldots, {f^{n-1}(\ket{\Psi_0})})\begin{pmatrix}
                1\\
                t_i/1!\\
                t_i^2/2! \\
                : \\
                t_i^{n-1}/({n-1})!
            \end{pmatrix}.
        \end{align}
        Writing the vectors as column vectors in a matrix results in 
        \begin{align}
             &({h_0^{(n)}}(\ket{\Psi_0}), \ldots, {h_{n-1}^{(n)}}(\ket{\Psi_0})) \nonumber \\
            &= ({f^0(\ket{\Psi_0})},{f^1(\ket{\Psi_0})}, \ldots, {f^{n-1}(\ket{\Psi_0})}) \Theta
        \end{align}
        with
        \begin{align}
            \Theta &= \begin{pmatrix}
                1 & 1 & \ldots & 1 \\
                t_1/1! & t_2/1! & \ldots & t_{n-1}/1! \\
                t_1^2/2! & t_2^2/2! & \ldots & t_{n-1}^2/2! \\
                : & : & : & : \\
                t_1^{n-1}/(n-1)! & t_2^{n-1}/(n-1)! & \ldots & t_{n-1}^{n-1}/(n-1)!
            \end{pmatrix}.
        \end{align}
        For $i\neq j$, $t_i \neq t_j$ is given. It follows that all column vectors of $\Theta$, called Vandermonde matrix, are linearly independent, and therefore $\Theta$ is invertible. The vectors of $\mathrm{K}_m$ can then be represented with the inverse $\Theta^{-1}$ as
        \begin{align}
            ({f^0(\ket{\Psi_0})}, \ldots, {f^{n-1}(\ket{\Psi_0})}) & = ({h_0^{(n)}}(\ket{\Psi_0}), \ldots, {h_{n-1}^{(n)}}(\ket{\Psi_0})) \Theta^{-1}.
        \end{align}
        As all vectors ${f^0(\ket{\Psi_0})}, \ldots, {f^{n-1}(\ket{\Psi_0})}$ of $\mathrm{K}_n$ can be represented by the basis vectors ${h_0^{(n)}}(\ket{\Psi_0}), \ldots, {h_{n-1}^{(n)}}(\ket{\Psi_0})$ of $\mathrm{H}_n^n$, it follows that $\mathrm{H}_n^n = \mathrm{K}_n$ holds true. 
    \end{proof}
\end{theorem}

\begin{theorem}[Approximation of Krylov Space]
\label{theorem:2_2_main_proof}
    Assume everything like in \cref{theorem:2time_lim}. Then for a large $N_L \in \mathbb{N}$
    \begin{align}
        G_{N_L} = \mathrm{Span} (\ket{g_0},..,\ket{g_{{N_L}-1}}), 
    \end{align}
    with some $\ket{g_i}=\exp (-iHt_i)\ket{\Psi_0}$, it holds that 
    \begin{align}
        \mathrm{H}_{N_L}^{N_L} \approx \mathrm{G}_{N_L}^{N_L}
    \end{align}
    From this it follows then that $\mathrm{G}_{N_L} \approx \mathrm{K}_{N_L}$ holds as well and finally $\mathrm{G}_{m} \approx \mathrm{K}_{m}$. 
    \begin{proof}
    We consider a $ N_L\gg m$. This leads to
        \begin{align}
            \mathrm{K}_m &= \mathrm{K}_{N_L} = \mathrm{H}_{N_L}^{N_L} \nonumber \\
            &= \mathrm{Span}\Big(\sum_{j=0}^{{N_L}-1}{f^j(\ket{\Psi_0})}\frac{t_0^j}{j!}, \ldots, \sum_{j=0}^{{N_L}-1}{f^j(\ket{\Psi_0})}\frac{t_{{N_L}-1}^j}{j!}\Big) \label{eq:K=H}
        \end{align}
    For any $t>0$ and sufficiently large $N_L$ the series terms $t^{N_L}/N_L!\rightarrow 0$, such that the approximation
    \begin{align}
        \sum_{j=0}^{{N_L}-1}{f^j(\ket{\Psi_0})}\frac{t^j}{j!} \approx \sum_{j=0}^\infty {f^j(\ket{\Psi_0})}\frac{t^j}{j!}
    \end{align}
    can be used. Inserting this into \cref{eq:K=H} results in
    \begin{align}
         H_{N_L}^{N_L} &=\mathrm{Span}\Big(\sum_{j=0}^{{N_L}-1}{f^j(\ket{\Psi_0})}\frac{t_0^j}{j!}, \ldots, \sum_{j=0}^{{N_L}-1}{f^j(\ket{\Psi_0})}\frac{t_{{N_L}-1}^j}{j!}\Big) \nonumber \\
         &\approx \mathrm{Span}\Big(\sum_{j=0}^\infty {f^j(\ket{\Psi_0})}\frac{t_0^j}{j!}, \ldots, \sum_{j=0}^{\infty}{f^j(\ket{\Psi_0})}\frac{t_{{N_L}-1}^j}{j!} \Big) \nonumber \\
         \Leftrightarrow  &K_{N_L}=H_{N_L}^{N_L} \approx G_{N_L}.
    \end{align}
    As $\mathrm{K}_{N_L}$ is a Krylov space, there exists a grade $m$ such that $\mathrm{K}_{p>m}=\mathrm{K}_m$ holds true. Since there are only $m$ linearly independent vectors in $\mathrm{K}_m$, it follows that $\mathrm{H}_{N_L}^{N_L}$ only has $m$ linearly independent vectors. For very large ${N_L}$, we conclude that $\mathrm{G}_{N_L}$ also has $m$ linearly independent vectors. 
    The $m$ linearly independent vectors $\ket{g_i}$ will be addressed with the subscripts $i_0, i_1, \ldots, i_{m-1}$ as $\ket{g_{i_0}}, \ket{g_{i_1}}, \ldots, \ket{g_{i_{m-1}}}\in \{\ket{g_0}, \ket{g_1}, \ket{g_2}, \ldots \}$. This leads to the result
    \begin{align}
        \mathrm{G}_m = \mathrm{Span}\Big(\ket{g_{i_0}}, \ket{g_{i_1}}, \ldots, \ket{g_{i_{m-1}}}\Big) \approx \mathrm{K}_m.
    \end{align}
\end{proof}
\end{theorem}
The property of $\mathrm{H}_{N_L}^{N_L}=\mathrm{K}_{N_L}$ was shown in the prior example for any ${N_L}\in\mathbb{N}$.  In the case of ${N_L}=\infty$ the approximation would reduce to $\mathrm{K}_{N_L}=\mathrm{G}_{N_L}$, which implies $\mathrm{G}_{m}=\mathrm{K}_m$. In our simulations we observed not only the approximation $\mathrm{G}_m\approx \mathrm{K}_m$ but the equality of these two spaces, i.e. $\mathrm{G}_m=\mathrm{K}_m$. 
The benefit of using the vectors $\ket{g_i}$ as a basis instead of the vectors ${f^i(\ket{\Psi_0})}$ is that these vectors are normalized as $|\ket{g_i}|=|\exp(-iHt_i)\ket{\Psi_0}|=|\ket{\Psi_0}|=1$. $\ket{g_i}$ are therefor robust to numerical rounding when checking linear independence. Additionally, with the exception of a global phase $\ket{g_i} \rightarrow \exp(i\alpha)\ket{g_i}$, the vectors $\ket{g_i}$ can be reconstructed through measurements. \\
Later, we will perform the Gram-Schmidt algorithm on $\mathrm{G}_m$ to obtain the orthonormal basis $\big\{\ket{w_i}\big\}_i$, such that
\begin{align}
    \mathrm{G}_m &= \mathrm{Span}\Big(\ket{g_{0}}, \ldots, \ket{g_{{m-1}}}\Big) 
    = \mathrm{Span}\Big(\ket{w_{0}}, \ldots, \ket{w_{{m-1}}}\Big)
\end{align}
holds.
The following two lemmas will add to \cref{theorem:2time_lim} to show that the basis can be computed through quantum mechanical measurements. 
\begin{lemma}
    \label{lemma:1globalPhase}
    The vector space $\mathrm{G}_m$ is invariant under a phase $\alpha\in \mathbb{R}$. 
    \begin{proof}
    \begin{align}
        e^{i\alpha } \mathrm{K}_m &= e^{i\alpha}\text{Span}(\ket{g_0},..,\ket{g_{m-1}}) \nonumber \\
        &= \text{Span}(e^{i\alpha}\ket{g_0},..,e^{i\alpha}\ket{g_{m-1}})  =\mathrm{K}_m
    \end{align}
    In the last step, we have used that $z=e^{i\alpha}\neq 0$ is a complex number. This can be extended to different complex phases $\alpha_a$. 
    \begin{align}
        \mathrm{K}_m &= \text{Span}(e^{i\alpha_0}\ket{g_0},..,e^{i\alpha_{m-1}}\ket{g_{m-1}}) 
        =\mathrm{K}_m
    \end{align}
    \label{proof:globalPhase}
    \end{proof}
\end{lemma}

\begin{lemma}
\label{lemma:equidistantTimes}
    Consider a quantum system with Hamiltonian $H$ having eigenvectors $\ket{\phi_i}$ and corresponding eigenvalues $\epsilon_i$ given by
    \begin{align}
        H\ket{\phi_i} = \epsilon_i\ket{\phi_i}.
    \end{align} 
    Let $\ket{\Psi_0}$ be a non-trivial initial state, specifically a superposition of at least two eigenstates $\ket{\phi_n}$ and $\ket{\phi_m}$ with distinct eigenvalues $\epsilon_n \neq \epsilon_m$. Additionally, assume two distinct times $t_1$ and $t_2$, both less than the system's period $T_P$. Then, it follows that the two time-evolved states $\ket{g_1} = \exp(-iHt_1)\ket{\Psi_0}$ and $\ket{g_2} = \exp(-iHt_2)\ket{\Psi_0}$ are linearly independent.
    \begin{proof}
    The two states $\ket{g_1}$ and $\ket{g_2}$ can be represented as   
    \begin{align}
        \ket{g_1} &= e^{-iHt_1}\ket{\Psi_0} = \sum_i e^{-i\epsilon_it_1}\ket{\phi_i}\bra{\phi_i}\ket{\Psi_0}\\
        \ket{g_2} &= e^{-iHt_2}\ket{\Psi_0} = \sum_i e^{-i\epsilon_it_2}\ket{\phi_i}\bra{\phi_i}\ket{\Psi_0}.
    \end{align}
    Substituting $\alpha_i = \bra{\phi_i} \ket{\Psi_0}$ leads to
    \begin{align}
        \ket{g_1} &= e^{-iHt_1}\ket{\Psi_0} =\sum_i \alpha_i e^{-i\epsilon_it_1}\ket{\phi_i} 
        = (\ket{\phi_1}, \ket{\phi_2}, .., \ket{\phi_N})\begin{pmatrix}
            \alpha_1e^{i\epsilon_1t_1}\\
            \alpha_2e^{i\epsilon_2t_1}\\
            :\\
            \alpha_Ne^{i\epsilon_Nt_1}
        \end{pmatrix}\\
        \ket{g_2} &= e^{-iHt_2}\ket{\Psi_0} =\sum_i \alpha_i e^{-i\epsilon_it_2}\ket{\phi_i} 
        = (\ket{\phi_1}, \ket{\phi_2}, .., \ket{\phi_N})\begin{pmatrix}
            \alpha_1e^{i\epsilon_1t_2}\\
            \alpha_2e^{i\epsilon_2t_2}\\
            :\\
            \alpha_Ne^{i\epsilon_Nt_2}
        \end{pmatrix}\\
    \end{align}
    The matrix representation $\mathbf{a}_i$ of the vector $\ket{g_i}$ in regards to the eigenstates $\Big\{\ket{\phi_i}\Big\}$ is given by
    \begin{align}
        \mathbf{a}_i:= \begin{pmatrix}
            \alpha_1e^{i\epsilon_1t_i}\\
            \alpha_2e^{i\epsilon_2t_i}\\
            :\\
            \alpha_Ne^{i\epsilon_Nt_i}
        \end{pmatrix}.
    \end{align}
    Given the condition that there exist at least two distinct eigenvalues $\epsilon_n \neq \epsilon_m$ with non-zero coefficients $\alpha_n$ and $\alpha_m$ implies that the basis representations $\mathbf{a}_1$ and $\mathbf{a}_2$ are independent. Due to this independence and the orthonormality of the eigenstates, it follows that the states $\ket{g_1}$ and $\ket{g_2}$ are linearly independent.
    Since any two states $\ket{g_i}$ and $\ket{g_j}$ are linearly independent for any distinct $t_i \neq t_j$, and as there exist a total of $m$ linearly independent vectors $\ket{g_i}$ (as described in \cref{theorem:2time_lim}), it follows that for any $t_0 < t_1 < t_2 < \ldots < t_{m-1} = T_G  < T_P$, $m$ linearly independent vectors are obtained, represented by $\ket{g_i} = \exp(-iHt_i)\ket{\Psi_0}$.
    In the case of an equidistant time axis, meaning $t_j = (j+1)\tau/N$ for some $\tau<T_P$ and $i<m$, and with $\ket{g_j}= \exp(-iHt_j)\ket{\Psi_0}$, this results in
    \begin{align}
        \mathrm{G}_m = \text{Span}\Big(\ket{g_0},\ket{g_1},..,\ket{g_{m-1}}\Big) \approx \mathrm{K}_m.
    \end{align}
    \end{proof}
\end{lemma}
The space $\mathrm{G}_m$ comprises of states with the common initial state $\ket{\Psi_0}$ and varying evolution times $t_i$. Notably, because of the invariance under a global phase (\cref{lemma:1globalPhase}) $\mathrm{G}_m$ 
 can be computed through quantum measurements. 
 
\cref{theorem:3Eigenvalue_EQ} will discuss the similarities between $\mathrm{G_m}$ and a basis representation through the eigenstates of the system. We show that through a combination of the eigenstates of the Hamiltonian the time-evolution can be represented through $d$ superpositions of eigenstates of the system, where $d$ is the number of pairwise distinct eigenvalues. This will help us to understand the Krylov space and its meaning. If both spaces have the minimum number of basis states, then the dimension of these two spaces should be equivalent. This will further be numerically verified in \cref{sec:num_ver}.

\begin{theorem}
\label{theorem:3Eigenvalue_EQ}
    Consider a Hamiltonian $H \in \mathbb{C}^{N \times N}$ with $H = H^\dagger$, possessing $d$ pairwise distinct eigenvalues, and satisfying the eigenvalue equation
    \begin{align}
        H\ket{\phi_j} = \epsilon_j\ket{\phi_j}.
    \end{align}
    Then, it follows that the dimension (grade) of the Krylov space $\mathrm{K}_m$ is equivalent to the count of pairwise distinct eigenvalues $\epsilon_j$.
    \begin{proof}
    With $\alpha_j=\bra{\phi_j}\ket{\Psi_0}$, it follows:
        \begin{align}
            \ket{\Psi(t)} &= e^{-iHt}\ket{\Psi_0} = \sum_j e^{-i\epsilon_j t}\ket{\phi_j}\bra{\phi_j}\ket{\Psi_0} = \sum_j e^{-i\epsilon_j t}\alpha_j  \ket{\phi_j}
        \end{align}
        Since $H$ is an $N\times N$ matrix, there exist $N$ eigenvalues, where only $d$ are pairwise distinct. The $d$ pairwise distinct eigenvalues are $\epsilon_1, \ldots, \epsilon_d$. As $H$ is Hermitian, it follows that the eigenstates $\ket{\phi_j}$ form an orthogonal basis of $\mathbb{C}^N$. Assume that there are $j_i$ eigenvectors with the same eigenvalue $\epsilon_i$. Introduce the index sets $J_1=\{1, \ldots, j_1\}$ and $J_k=\{j_{k-1}+1, \ldots, j_k\}$ with $k\in \{2, \ldots, d\}$. The index sets group eigenstates $\ket{\phi_i}$, so that all eigenstates $\ket{\phi_a}$, with $a\in J_k$, have the same corresponding eigenvalue $\epsilon_k$. 
        This can be used to rewrite the time evolution as:
        \begin{align}
            \ket{\Phi(t)} &= \sum_j e^{-i\epsilon_j t}\alpha_j \ket{\phi_j}
            = \sum_{p=1}^{d}\sum_{j\in J_p}e^{-i\epsilon_pt}\alpha_j\ket{\phi_j}
            = \sum_{p=1}^{d}e^{-i\epsilon_pt} \sum_{j\in J_p}\alpha_j\ket{\phi_j}
        \end{align}
        The summation over the index sets $J_p$ is independent of time $t$ and can therefore be substituted with vectors $\ket{\xi_p} = \sum_{j\in J_p}\ket{\Phi_i}$, which results in
        \begin{align}
            \ket{\Phi(t)} = \sum_{p=1}^d e^{-i\epsilon_p t}\ket{\xi_p}.
        \end{align}
        Such that for all $t\in \mathbb{R}$ the following holds:
        \begin{align}
            \ket{\Phi(t)}\in \text{Span}\Big(\xi_1, \xi_2, \ldots, \xi_d \Big) := X_d.
        \end{align}
        The space $X_d$ consists of a superposition of $d$ linearly independent vectors $\ket{\xi_1},..,\ket{\xi_d}$. For any time any two vectors are linearly independent, which implies that $X_d$ is yet another minimal basis of the time evolution. 
        Let $\mathrm{K}_{m}$ be the Krylov space for $H$ initiated with the state $\ket{\Psi_0}$, where $m$ denotes its grade and is therefor a minimal basis. 
        Both spaces $X_d$ and $\mathrm{K}_m$ consist of the smallest number of vectors needed to represent any time-evolved state $\ket{\Phi(t)}$ with a initial condition $\ket{\Psi_0}$ in regards to the Hamiltonian $H$. Therefore, it follows that $d=m$ must hold true. 
    \end{proof}
\end{theorem}
Up until now, we have shown in \cref{theorem:2_2_main_proof} that time-evolved states exist, which span the same space as $\mathrm{K}_m$. We further showed the phase invariance (\cref{lemma:1globalPhase}) and that the dimension is equal to the number of pairwise distinct eigenvalues(\cref{theorem:3Eigenvalue_EQ}).
The next chapter will verify these results for simple systems. 

\section{Numerical verification and effective dimension}
\label{sec:num_ver}
To check whether $\mathrm{G}_m$ fully captures the evolution of a state under the Schrödinger equation, we will check if any time-evolved state can be reconstructed by the basis of $\mathrm{G}_m$ without error. Furthermore, we want to make sure that the dimension of $\mathrm{G}_m$ is equal to the number of pairwise distinct eigenvalues $d$. \\
First, consider a Hamiltonian $H \in \mathbb{C}^{N \times N}$ along with $N_S$ randomly chosen initial states $\ket{\Psi_1}, \ket{\Psi_2}, ..., \ket{\Psi_{N_S}}$. Select a timescale \( T_G < T_P \) for the construction of the space \( \mathrm{G} \), where \( T_P \) represents the period of the time evolution operator, i.e., \( \exp(-iH(T_P + t)) = \exp(-iHt) \) holds.
Proceed with a discretization $0=t_0<t_1<t_2<...<t_{N-1}=T_G<T_P$ and compute the vectors $\ket{g_i} = \exp(-iHt_i)\ket{\Psi_x}$ for a given starting state $\ket{\Psi_x}$, where $x=1, ..., N_S$. The matrix $A_m = (\ket{g_0}, \ket{g_1}, ..., \ket{g_{m-1}})$ is constructed iteratively, incrementing $k$ until $rank(A_{m-1})=rank(A_m)$ is reached for some $k=m$. Here, $m$ denotes the dimension of the vector space $\mathrm{K}_m$ and, by approximation, the dimension of $\mathrm{K}_m(-iH, \ket{\Psi_x})$ or the grade of $\ket{\Psi_x}$ in regards to $f:=-iH$. After forming $\mathrm{G}_m = Span\Big(\ket{g_0}, \ket{g_1}, ..., \ket{g_{m-1}}\Big)$, the Gram-Schmidt algorithm is applied to obtain orthonormalized vectors $\ket{w_0}, \ket{w_1}, ..., \ket{w_{m-1}}$, which results in
\begin{align}
    \mathrm{G}_m &= Span\Big(\ket{g_0}, ..., \ket{g_{m-1}}\Big) = Span\Big(\ket{w_0}, ..., \ket{w_{m-1}}\Big).
\end{align}
Once the vectors $\ket{w_i}$ are computed, $N_T$ random times $\tau_1, \tau_2, ..., \tau_{N_T}$ are chosen, and states evolved by the times $\ket{\Psi(\tau_k)} = \exp(-iH\tau_k)\ket{\Psi_x}$ are calculated. As per \cref{theorem:1EvolKrylov} $\ket{\Psi(\tau)} \in \mathrm{K}_m$ is true for all $\tau \in \mathbb{R}$. Furthermore, \cref{theorem:2time_lim} implies that $\mathrm{G}_m \approx \mathrm{K}_m$ is a valid approximation, which will be verified by demonstrating that for any time $\tau_k$, $\ket{\Psi(\tau_k)} \in \mathrm{G}_m$ holds true.
The $l$-th approximation $\ket{u_l(\tau_k)}$ of $\ket{\Psi(\tau_k)}$ concerning the starting state $\ket{\Psi_x}$ is given by
\begin{align}
    \ket{u_l(\tau_k)} = \sum_{j=0}^{l-1} \bra{w_j}\ket{\Psi(\tau_k)} \ket{w_j}.
\end{align}
The error in the $l$-th approximation of a time-evolved state $\ket{\Psi(\tau_k)}$ with the initial vector $\ket{\Psi_x}$ is quantified by
\begin{align}
    r(\tau_k, l, \ket{\Psi_x}) = \Big| \ket{u_l(\tau_k)} - \ket{\Psi(\tau_k)} \Big|.
\end{align}
If $\mathrm{G}_m=\mathrm{K}_m$ holds true, then $r(\tau_k, m,\ket{\Psi_x}) = 0$ is expected. For a statistical analysis, multiple times $\tau_k$ are considered, and the average over these times is calculated as $r(l, \ket{\Psi_x})$.
\begin{align}
    r(l, \ket{\Psi_x}) = \frac{1}{N_T}\sum_{k=1}^{N_T} r(\tau_k, l, \ket{\Psi_x}).
\end{align}
Moreover, $N_S$ initial conditions $\ket{\Psi_x}$, where $x=1, ..., N_S$, are provided. After averaging over all possible initial conditions, the error of the $l$-th approximation is expressed as
\begin{align}
    r(l) = \sum_{x=1}^{N_S} r(l, \ket{\Psi_x}).
\end{align}
If $\mathrm{G}_m=\mathrm{K}_m$ is confirmed, where $m$ is chosen such that $r(m)=0$ for all $\ket{\Psi_x}$ and all $\tau_k$, then the approximation is deemed valid. The number of distinct eigenvalues $d$ is expected to be equal to $m$, i.e., $m=d$, for all Hamiltonians.
In an $N_Q$-qubit system, the $N_Q$-qubit Pauli matrices $P_i\in \{X_i,Y_i,Z_i\}$ are constructed as
\begin{align}
    P_i = \bigotimes_{j=1}^{i-1}I_2 \otimes \sigma_p \otimes \bigotimes_{j=i+1}^{N_Q} I_2,
\end{align}
where $p\in \{\sigma_x, \sigma_y, \sigma_z\}$, and $I_2$ represents the Pauli matrices and the 2-dimensional identity.

\begin{figure}[H]
    \hspace*{-0.5 cm}
    \centering
    \includegraphics[scale=1]{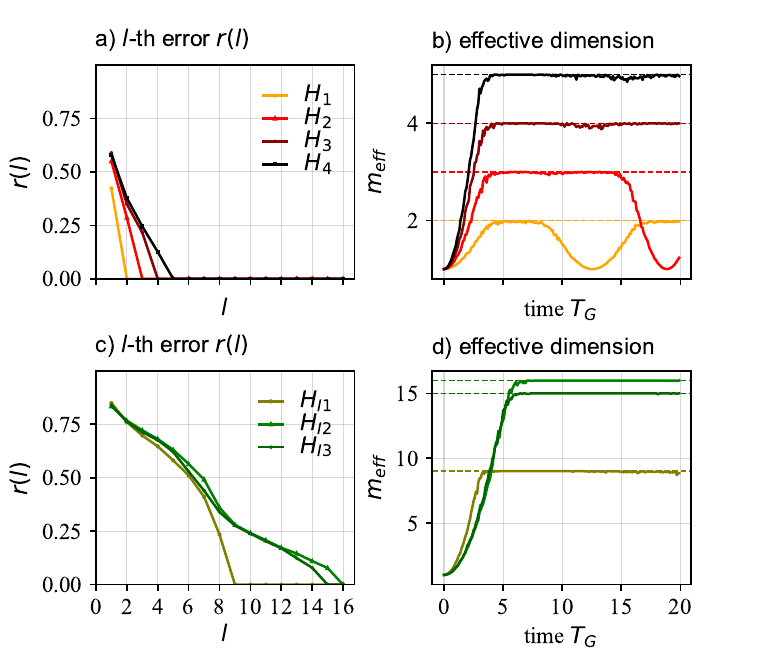}
    \caption[State reconstruction through Krylov space ]{The $l$th averaged reconstruction error in \textbf{a}) and \textbf{c}) for the simple Hamiltonians $H_1$ to $H_4$ and Ising Hamiltonians $H_{I1}$ to $H_{I3}$ and the effective Krylov dimension in \textbf{b}) and \textbf{d}) for the same Hamiltonians. }
    \label{fig:1_SC}
\end{figure}

In this study, we conduct an analysis of various four-qubit Hamiltonians. Initially, attention is given to simple Hamiltonians denoted as $H_1, H_2, H_3$, and $H_4$, which consist of rotation in the x-direction of the first, first two, first three, and all four qubits respectively. 
\begin{align}
    H_k = \sum_{i=1}^k 0.5 X_k
\end{align}
Subsequently, the analysis expands to include Ising Hamiltonian $H_{I1}, H_{I2}$ and $H_{I3}$ with different inter-spin couplings $J_{ij}$.
\begin{align}
    H_I = \sum_{i=1,j>i}^4 J_{ij}X_iX_j + 0.5\sum_{i=0}^4 Z_i
\end{align}
Further information on the couplings $J_{ij}$ is listed in \cref{sec:appendix}.
Our objective is to assess whether the error $r(m)=0$ and if $m=d$ holds. \cref{table:grade_eigs} shows the calculated grade $m$ of the Krylov space and the number of pairwise distinct eigenvalues for all Hamiltonian. We observe in each case the equivalence between the two values, which gives numerical verification of \cref{theorem:3Eigenvalue_EQ}.
\begin{table}[H]
\centering
\begin{tabular}{|l|l|l|l|l|l|l|l|}
\hline
                                   & $H_1$ & $H_2$ & $H_3$ & $H_4$ & $H_{I1}$ & $H_{I2}$ & $H_{I3}$ \\ \hline
Grade of $\mathrm{G}_m$, $m$            & 2     & 3     & 4     & 5     & 9        & 16       & 15       \\ \hline
\#eigenvalues, $d$ & 2     & 3     & 4     & 5     & 9        & 16       & 15       \\ \hline
\end{tabular}
\caption{The grades $m$ of the different quantum systems and the number of pairwise distinct eigenvalues $d$. For each system equality is observed.}
\label{table:grade_eigs}
\end{table}
\cref{fig:1_SC}.a) illustrates the averaged error of the reconstruction error $r(l)$ of the simple Hamiltonians $H_i$. In each scenario, reconstruction is evident when $l=m$, as indicated by $r(m)=0$ for each Hamiltonian. By increasing the complexity of the Hamiltonian from $H_1$ to $H_2$ and others, we increase the number of pairwise distinct eigenvalues and the dimension of the Krylov space, as shown in \cref{table:grade_eigs}.
Next, we compute the Krylov spaces and calculate the errors for some realizations of the Ising Hamiltonian, denoted as $H_{I1}$, $H_{I2}$, and $H_{I3}$ and show the reconstruction error $r(l)$ in \cref{fig:1_SC}.c). Even though the Hamiltonian $H_{I3}$ seems more complex compared to $H_{I2}$, the Krylov dimension and the number of pairwise distinct eigenvalues are smaller. Here as well, the correct reconstruction of all states by the calculated spaces $G_m$ can be observed.

These numerical results emphasize that the defined space $\mathrm{G}_{m}$ forms a robust basis resilient to numerical errors, a quality that can be further verified through quantum-mechanical measurements.\\
Next, an effective space dimension will be constructed to describe the space $G_m$. For the effective dimension, we will look at states after they are evolved by a time $T_G<T_P$. We compute states $\ket{g_i}=e^{-iH\theta_i}\ket{\Psi_0}$ with evolution times $\theta_i=(i+1)T_G/m$ where $i=1,..,m$, and $m$ is the Krylov space dimension of the corresponding system. By \cref{lemma:equidistantTimes}, we know that all $\ket{g_i}$ are linearly independent in the mathematical sense. If two vectors are very close to each other, i.e., $\ket{g_i}\approx \ket{g_{i+1}}$, we require an effective dimension to consider the second vector. To obtain an effective dimension, we calculate the square of the fidelity $F$ 
\begin{align}
    \lambda_{i} =  F\Big(\ket{g_i}\bra{g_i}, \ket{g_{i+1}}\bra{g_{i+1}}\Big)^2 =\abs{\bra{g_i}\ket{g_{i+1}}}^2.
\end{align}
The effective dimension $m_{eff}$ for $\lambda = 1/\sqrt{2}$ is given by
\begin{align}
    m_{{eff}_i} &= \begin{cases}
        1 &\text{if } \lambda_i < \lambda\\
        1-\frac{1}{1-\lambda}\cdot(\lambda_i-\lambda)  &\text{if } \lambda_i \geq \lambda
    \end{cases} \nonumber \\
    m_{eff} &= 1 +\sum_{i=1}^{m-1}m_{{eff}_i}
\end{align}
The first vector increases the dimension by $1$. Afterwards we perform the interpolation between two states $\ket{g_i}$. $\lambda_i=1$ implies that the two vectors are linearly dependent and therefore numerically dependent, resulting in $m_{{eff}_i}=0$. $\lambda_i<\lambda$ implies that the two vectors are different enough to consider full numerical independence, and the effective dimension is then increased by $m_{{eff}_i}=1$. For $\lambda<\lambda_i\leq 1$ the effective dimension is interpolated. The measure $m_{{eff}_i}$ is upper-bounded by the Krylov space dimension $m_{{eff}_i}\leq m$ and is shown for different systems in \cref{fig:1_SC}.b) and \cref{fig:1_SC}.d). In the first row, we see the effective dimension for the four Hamiltonians $H_1, H_2, H_3$, and $H_4$. The effective dimension of $H_1$ (orange line) starts at one and then increases to the number of pairwise distinct eigenvalues, indicated by an orange horizontal line. Interestingly, we observe a decrease in effective dimension for $T=4\pi$. This can be explained by the $2\pi$ periodicity of the time evolution under $H_1$ as $\exp(-i2\pi X_1/2)=\exp(-i X_1/2)=-\exp(-i4\pi X_1/2)$. However, because the dimension of the Krylov space is $m=2$, we sample two time-evolved states at time $\theta_1=T_G/2$ and $\theta_2=T_G$. For $\theta=4\pi$, both are multiples of $2\pi$, which leads to the two vectors being $\ket{g_1}=-\ket{g_2}$. $\ket{g_2}$ being a scalar multiple of $\ket{g_1}$ implies linear dependence, which in turn reduces the effective dimension $m_{\text{eff}}$.
$H_2$ shows a small effective dimension for small times and increases for larger $T_G$. The vectors of $G_m$ are sampled at times $\theta_1=T_G/3, \theta_2=2T_G/3$, and $\theta_3=T_G$. The evolved states are linearly dependent for $T_G=3\cdot 2\pi$, as can be seen by the reduced effective dimension of the red curve for $T_G=3\cdot 2\pi$.
The effective dimension of $H_3$ and $H_4$ follows the same trend, where it starts at one and then increases towards the number of pairwise distinct eigenvalues, where it saturates.
In the second row, the effective dimension of the Ising Hamiltonians can be seen. An increase for larger times and then a saturation towards the number of pairwise distinct eigenvalues can be observed. \\

\section{Measurable Spread Complexity}\label{sec5:msc}
To better understand the utility of the constructed space $G$ and its basis $\ket{w_i}$, we want to compare the spread of these two different spaces $K$ and $G$ to each other. Assume a time-evolved state $\ket{\Psi (t)}$, represented in the Krylov basis $\ket{k_i}$ used in the usual computation for spread complexity and in the representation of our developed basis $\ket{w_i}$:
\begin{align}
    \ket{\Psi(t)}&=\sum_{n=0}^{m-1}\alpha_n(t)\ket{k_n} \nonumber \\
    &=\sum_{n=0}^{m-1}\beta_n(t)\ket{w_n}
\end{align}
We define the two complexities $C_H$ and $C_e$ as:
\begin{align}
    C_H(t) = \sum_{n=0}^{m-1} (n+1) \abs{\alpha_{n}(t)}^2 \\
    C_e(t) = \sum_{n=0}^{m-1} (n+1) \abs{\beta_{n}(t)}^2.
\end{align}
Note that the spread $C_e$ is computed through a quantum-mechanically measurable basis.

\begin{figure}[h]
    \hspace*{-0.5 cm}
    \centering
    \includegraphics[scale=1]{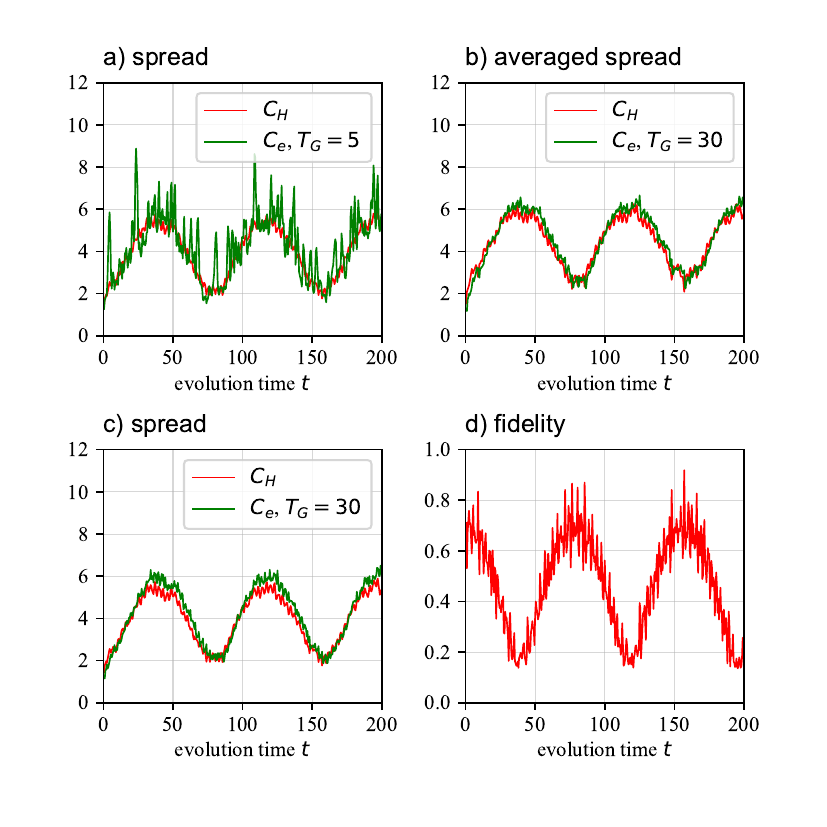}
    \caption[Comparison of measurable spread complexity]{The spread complexities $C_H$ and $C_e$ for $H_{I3}$ for different times in construction of the basis $\ket{g_i}$ in \textbf{a}) for $T=5$ and in \textbf{c}) for $T=30$ with one initial condition. The averaged spread complexity $C_H$ and $C_e$ over one hundred initial conditions in \textbf{b}) and the fidelity averaged over one hundred initial conditions in \textbf{d}).}
    \label{fig:2_sc}
\end{figure}

\cref{fig:2_sc}.a) shows the result for the measurable spread complexity $C_e$, where a time $T_G=5$ was used to construct $G$ and its basis $\ket{w_i}$. $C_e$ seems to consist of a term, which is similar to $C_H$ and a noisy oscillation. We have shown this result for one initial condition, but this is observable for any initial condition with $T_G=5$. We can make use of the effective dimension at $T_G=5$ in \cref{fig:1_SC}.c) for $H_{I3}$ to better understand this behavior. At $T=5$, it can be seen that the maximum of the effective dimension is not yet reached for $\lambda=0.707$, which indicates that the space $G$ has not fully captured the system dynamics. \cref{fig:2_sc}.c) shows the same initial condition with $T_G=30$ for the construction of $G$. Here, the behavior between $C_H$ and $C_e$ is almost identical, where the classical spread $C_H$ seems to be slightly lower than $C_e$. Averaging both spreads over one hundred initial conditions leads to the plot in \cref{fig:2_sc}.b), which results in almost identical spread, thus enabling the computation of spread complexity through measurements. 

Another interesting aspect of the results is the decrease for times that are multiples of $t\approx 80$ for $C_H$ and $C_e$. Typically, spread complexity starts at one and then increases as the state evolves through the system dynamics. To further understand this, we compute the fidelity $F_d$ between the initial state $\ket{\Psi(0)}$ and the state evolved by the time $t$:
\begin{align}
    F_d = \abs{\bra{\Psi(0)}\ket{\Psi(t)}}
\end{align}
We observe that as the fidelity increases, the spread complexity decreases, and when the fidelity is maximal, the spread is minimal. This can be observed in this very simple system due to its small number of eigenvalues, which lead to semi-periodic behavior for small evolution times.

\section{Discussion and Conclusion}
In this work, we discuss Krylov spaces for pure states, building upon prior research on operator and state complexity as explored in \cite{PAR19, BAL22, ALI23}. The basis vectors of the already researched vector space $\mathrm{K}$ require a large number of matrix multiplications to compute the powers $H^i$. Furthermore, knowledge about the Hamiltonian is necessary. Since $H$ is a hermitian matrix, the experimental computation of $H^i$ is currently unfeasible, requiring classical machine computation. While manageable for small systems, this becomes challenging for large quantum systems.

These challenges find resolution with the introduction of the space $\mathrm{G}$, where each basis vector consists of a state evolved over time under the corresponding Hamiltonian, ensuring normalization. We first show the equivalence between the Krylov space $\mathrm{K}$ and the state space $\mathrm{G}$ (see \cref{theorem:2time_lim}). \cref{theorem:3Eigenvalue_EQ} shows that the dimensions of $\mathrm{K}$ and $\mathrm{G}$ should be $d$, where $d$ represents the number of pairwise distinct eigenvalues of the Hamiltonian $H$.

Finally, we demonstrate the invariance of the vector space $\mathrm{G}$ under a global phase, which facilitates the computation of basis vectors $\ket{g_j}$ through quantum mechanical measurements. This property allows for the utilization of this expressivity measure in large quantum mechanical systems and aids in determining the number of pairwise distinct eigenvalues in a quantum mechanical experiment, as illustrated in \cref{lemma:1globalPhase}. To assess the representation accuracy of $\mathrm{G}$, we simulate various Hamiltonians and compute the dimension $m$ of $\mathrm{G}$ along with the number of pairwise distinct eigenvalues $d$ of the Hamiltonian in \cref{sec:num_ver}. For all simulated systems, we observe state representation at $m=d$, as indicated by a vanishing reconstruction error $r(m=d)=0$. This demonstrates that the defined space $\mathrm{G}$ is the smallest basis capable of representing any time-evolved state. We further expand on the description of $\mathrm{G}$ to define an effective dimension that characterizes the expressivity of the system, which can be of interest in understanding different quantum systems and quantum machine learning protocols.

Lastly, we computed the spread complexity for our introduced basis and the basis constructed using the Lanczos algorithm and compared the results in \cref{sec5:msc}. We can see that the spread complexities exhibit the same behavior if the evolution time is chosen sufficiently large, i.e., where the effective dimension or the spread complexity already reaches a maximum. If this is not given, a noisy oscillation on $C_e$ is observed. These results show that the basis does not necessarily need to consist of different powers of the Hamiltonian to compute spread complexity and define an effective dimension.

It would be interesting to analyze the influence of different bases on spread complexity. Here, we restricted ourselves to finding a measurable basis showing the equivalence to the basis constructed through different powers. The benefit of the classical approach to the computation of spread complexity is that it is faster to simulate on a classical system. However, this approach requires knowledge of a Hamiltonian and classical computation. For quantum systems where only time-evolved states are known, our approach can be used to construct a space $\mathrm{G}$ in which all time-evolved states are included.

The space $\mathrm{G}$ enables the computation of the spread complexity $C_e$ and the effective dimension $m_{eff}$, which can be used to understand the evolution of quantum systems where knowledge of the Hamiltonian is missing. This is especially important in quantum reservoir computing and quantum machine learning, where expressivity measures hold significant importance. In quantum reservoir computing, our preliminary research has shown that the effective Krylov dimension follows a similar trend to the information processing capacity for larger evolution times \cite{DAM12, MAR23}. Using the effective Krylov dimension in quantum machine learning training would also be of great interest for variational quantum circuits \cite{SCH15a, MAT21, MCC16, SCH21}. Another research interest is to discuss measurable operator spaces and the influence of Krylov complexity \cite{PAR19}.

\newpage
\bibliographystyle{JHEP}
\bibliography{lit}

\providecommand{\href}[2]{#2}\begingroup\raggedright\begin{thebibliography}{10}

\bibitem{PAR19}
D.E.~Parker, X.~Cao, A.~Avdoshkin, T.~Scaffidi and E.~Altman, \emph{A universal operator growth hypothesis}, \href{https://doi.org/10.1103/physrevx.9.041017}{\emph{Phys. Rev. X} {\bfseries 9} (2019) 041017}.

\bibitem{BAL22}
V.~Balasubramanian, P.~Caputa, J.M.~Magan and Q.~Wu, \emph{Quantum chaos and the complexity of spread of states}, \href{https://doi.org/10.1103/physrevd.106.046007}{\emph{Phys. Rev. D} {\bfseries 106} (2022) 046007}.

\bibitem{ALI23}
M.~Alishahiha and S.~Banerjee, \emph{A universal approach to {Krylov} state and operator complexities}, \href{https://doi.org/10.21468/scipostphys.15.3.080}{\emph{SciPost Phys.} {\bfseries 15} (2023) 080}.

\bibitem{AFR23}
M.~Afrasiar, J.K.~Basak, B.~Dey and K.~Pal, \emph{Time evolution of spread complexity in quenched {Lipkin}–{Meshkov}–{Glick} model}, \href{https://doi.org/10.1088/1742-5468/ad0032}{\emph{J. Stat. Mech.} {\bfseries 2023} (2023) 103101}.

\bibitem{ANE24}
T.~Anegawa, N.~Iizuka and M.~Nishida, \emph{Krylov complexity as an order parameter for deconfinement phase transitions at large {N}}, \href{https://doi.org/10.1007/jhep04(2024)119}{\emph{J. High Energy Phys.} {\bfseries 2024} (2024) 119}.

\bibitem{BAE22}
S.~Baek, \emph{Krylov complexity in inverted harmonic oscillator}, \href{https://doi.org/10.48550/arxiv.2210.06815}{\emph{arXiv} (2022) }.

\bibitem{BAR19}
J.L.F.~Barb{\'{o}}n, E.~Rabinovici, R.~Shir and R.~Sinha, \emph{On the evolution of operator complexity beyond scrambling}, \href{https://doi.org/10.1007/jhep10(2019)264}{\emph{J. High Energy Phys.} {\bfseries 2019} (2019) 264}.

\bibitem{BHA22a}
A.~Bhattacharya, P.~Nandy, P.P.~Nath and H.~Sahu, \emph{Operator growth and {Krylov} construction in dissipative open quantum systems}, \href{https://doi.org/10.1007/jhep12(2022)081}{\emph{J. High Energy Phys.} {\bfseries 2022} (2022) 81}.

\bibitem{BHA22b}
B.~Bhattacharjee, X.~Cao, P.~Nandy and T.~Pathak, \emph{Krylov complexity in saddle-dominated scrambling}, \href{https://doi.org/10.1007/jhep05(2022)174}{\emph{J. High Energy Phys.} {\bfseries 2022} (2022) 174}.

\bibitem{BHA23}
A.~Bhattacharyya, D.~Ghosh and P.~Nandi, \emph{Operator growth and {Krylov} complexity in {Bose}-{Hubbard} model}, \href{https://doi.org/10.1007/jhep12(2023)112}{\emph{J. High Energy Phys.} {\bfseries 2023} (2023) 112}.

\bibitem{BHA24a}
B.~Bhattacharjee, P.~Nandy and T.~Pathak, \emph{Operator dynamics in {Lindbladian} {SYK}: a {Krylov} complexity perspective}, \href{https://doi.org/10.1007/jhep01(2024)094}{\emph{J. High Energy Phys.} {\bfseries 2024} (2024) 94}.

\bibitem{CAM23}
H.A.~Camargo, V.~Jahnke, K.~Kim and M.~Nishida, \emph{Krylov complexity in free and interacting scalar field theories with bounded power spectrum}, \href{https://doi.org/10.1007/jhep05(2023)226}{\emph{J. High Energy Phys.} {\bfseries 2023} (2023) 226}.

\bibitem{CAO21}
X.~Cao, \emph{A statistical mechanism for operator growth}, \href{https://doi.org/10.1088/1751-8121/abe77c}{\emph{J. Phys. A Math. Theor.} {\bfseries 54} (2021) 144001}.

\bibitem{CAP21}
P.~Caputa and S.~Datta, \emph{Operator growth in 2d {CFT}}, \href{https://doi.org/10.1007/jhep12(2021)188}{\emph{J. High Energy Phys.} {\bfseries 2021} (2021) 188}.

\bibitem{CHA23}
A.~Chattopadhyay, A.~Mitra and H.J.R.~Van~Zyl, \emph{Spread complexity as classical dilaton solutions}, \href{https://doi.org/10.1103/physrevd.108.025013}{\emph{Phys. Rev. D} {\bfseries 108} (2023) }.

\bibitem{DYM20}
A.~Dymarsky and A.~Gorsky, \emph{Quantum chaos as delocalization in {Krylov} space}, \href{https://doi.org/10.1103/physrevb.102.085137}{\emph{Phys. Rev. B} {\bfseries 102} (2020) 085137}.

\bibitem{DYM21}
A.~Dymarsky and M.~Smolkin, \emph{Krylov complexity in conformal field theory}, \href{https://doi.org/10.1103/physrevd.104.l081702}{\emph{Phys. Rev. D} {\bfseries 104} (2021) L081702}.

\bibitem{FAN22}
Z.~Fan, \emph{Universal relation for operator complexity}, \href{https://doi.org/10.1103/physreva.105.062210}{\emph{Phys. Rev. A} {\bfseries 105} (2022) 062210}.

\bibitem{GUO22A}
S.~Guo, \emph{Operator growth in {SU}(2) {Yang}-{Mills} theory}, {\emph{arXiv} (2022) }.

\bibitem{HAS23}
K.~Hashimoto, K.~Murata, N.~Tanahashi and R.~Watanabe, \emph{Krylov complexity and chaos in quantum mechanics}, \href{https://doi.org/10.1007/jhep11(2023)040}{\emph{J. High Energy Phys.} {\bfseries 2023} (2023) 40}.

\bibitem{HE22}
S.~He, P.H.C.~Lau, Z.~Xian and L.~Zhao, \emph{Quantum chaos, scrambling and operator growth in \$\$ {T}{\textbackslash}overline\{{T}\} \$\$deformed {SYK} models}, \href{https://doi.org/10.1007/jhep12(2022)070}{\emph{J. High Energy Phys.} {\bfseries 2022} (2022) 70}.

\bibitem{HEV22}
R.~Heveling, J.~Wang and J.~Gemmer, \emph{Numerically {Probing} the {Universal} {Operator} {Growth} {Hypothesis}}, \href{https://doi.org/10.1103/physreve.106.014152}{\emph{Phys. Rev. E} {\bfseries 106} (2022) 014152}.

\bibitem{IIZ23}
N.~Iizuka and M.~Nishida, \emph{Krylov complexity in the {IP} matrix model}, {\emph{arXiv} (2023) }.

\bibitem{JIA21}
S.~Jian, B.~Swingle and Z.~Xian, \emph{Complexity growth of operators in the {SYK} model and in {JT} gravity}, \href{https://doi.org/10.1007/jhep03(2021)014}{\emph{J. High Energy Phys.} {\bfseries 2021} (2021) 14}.

\bibitem{KIM22}
J.~Kim, J.~Murugan, J.~Olle and D.~Rosa, \emph{Operator delocalization in quantum networks}, \href{https://doi.org/10.1103/physreva.105.l010201}{\emph{Phys. Rev. A} {\bfseries 105} (2022) L010201}.

\bibitem{LI24}
X.~Li, Q.~Zhu, C.~Zhao, X.~Duan, B.~Zhao, X.~Zhang et~al., \emph{Higher-order granger reservoir computing: simultaneously achieving scalable complex structures inference and accurate dynamics prediction}, \href{https://doi.org/10.1038/s41467-024-46852-1}{\emph{Nat. Commun.} {\bfseries 15} (2024) 2506}.

\bibitem{LIU23b}
C.~Liu, H.~Tang and H.~Zhai, \emph{Krylov complexity in open quantum systems}, \href{https://doi.org/10.1103/physrevresearch.5.033085}{\emph{Phys. Rev. Res.} {\bfseries 5} (2023) 033085}.

\bibitem{MAG20}
J.M.~Magan and J.~Sim{\'{o}}n, \emph{On operator growth and emergent {Poincar{\'{e}}} symmetries}, \href{https://doi.org/10.1007/jhep05(2020)071}{\emph{J. High Energy Phys.} {\bfseries 2020} (2020) 71}.

\bibitem{MUC22}
W.~M{\"{u}}ck and Y.~Yang, \emph{Krylov complexity and orthogonal polynomials}, \href{https://doi.org/10.1016/j.nuclphysb.2022.115948}{\emph{Nucl. Phys. B.} {\bfseries 984} (2022) 115948}.

\bibitem{NIZ23}
A.A.~Nizami and A.W.~Shrestha, \emph{Krylov construction and complexity for driven quantum systems}, \href{https://doi.org/10.1103/physreve.108.054222}{\emph{Phys. Rev. E} {\bfseries 108} (2023) 054222}.

\bibitem{PAT22}
D.~Patramanis, \emph{Probing the entanglement of operator growth}, \href{https://doi.org/10.1093/ptep/ptac081}{\emph{Prog. Theor. Phys.} {\bfseries 2022} (2022) 063A01}.

\bibitem{RAB22}
E.~Rabinovici, A.~S{\'{a}}nchez-Garrido, R.~Shir and J.~Sonner, \emph{Krylov localization and suppression of complexity}, \href{https://doi.org/10.1007/jhep03(2022)211}{\emph{J. High Energy Phys.} {\bfseries 2022} (2022) 211}.

\bibitem{RAB22a}
E.~Rabinovici, A.~S{\'{a}}nchez-Garrido, R.~Shir and J.~Sonner, \emph{Krylov complexity from integrability to chaos}, \href{https://doi.org/10.1007/jhep07(2022)151}{\emph{J. High Energy Phys.} {\bfseries 2022} (2022) 151}.

\bibitem{VAS24}
M.J.~Vasli, K.B.~Velni, M.R.M.~Mozaffar, A.~Mollabashi and M.~Alishahiha, \emph{Krylov complexity in {Lifshitz}-type scalar field theories}, \href{https://doi.org/10.1140/epjc/s10052-024-12609-9}{\emph{EOJ C} {\bfseries 84} (2024) 235}.

\bibitem{CAP24}
P.~Caputa, H.~Jeong, S.~Yi, J.F.~Pedraza and L.~Qu, \emph{Krylov complexity of density matrix operators}, \href{https://doi.org/10.1007/jhep05(2024)337}{\emph{J. High Energy Phys.} {\bfseries 2024} (2024) 337}.

\bibitem{BAL22a}
F.~Ballar~Trigueros and C.~Lin, \emph{Krylov complexity of many-body localization: {Operator} localization in {Krylov} basis}, \href{https://doi.org/10.21468/scipostphys.13.2.037}{\emph{SciPost Phys.} {\bfseries 13} (2022) 037}.

\bibitem{BAL23}
V.~Balasubramanian, J.M.~Magan and Q.~Wu, \emph{Tridiagonalizing random matrices}, \href{https://doi.org/10.1103/physrevd.107.126001}{\emph{Phys. Rev. D} {\bfseries 107} (2023) 126001}.

\bibitem{CRA23}
B.~Craps, O.~Evnin and G.~Pascuzzi, \emph{A relation between {Krylov} and {Nielsen} complexity}, {\emph{arXiv} (2023) }.

\bibitem{AGU24}
S.E.~Aguilar-Gutierrez and A.~Rolph, \emph{Krylov complexity is not a measure of distance between states or operators}, {\emph{arXiv} (2024) }.

\bibitem{CAP22}
P.~Caputa and S.~Liu, \emph{Quantum complexity and topological phases of matter}, \href{https://doi.org/10.1103/physrevb.106.195125}{\emph{Phys. Rev. B} {\bfseries 106} (2022) 195125}.

\bibitem{CAP22a}
P.~Caputa, J.M.~Magan and D.~Patramanis, \emph{Geometry of {Krylov} complexity}, \href{https://doi.org/10.1103/physrevresearch.4.013041}{\emph{Phys. Rev. Research} {\bfseries 4} (2022) 013041}.

\bibitem{ERD23}
J.~Erdmenger, S.~Jian and Z.~Xian, \emph{Universal chaotic dynamics from {Krylov} space}, \href{https://doi.org/10.1007/jhep08(2023)176}{\emph{J. High Energy Phys.} {\bfseries 2023} (2023) 176}.

\bibitem{GIL23}
W.~Gilpin, \emph{Model scale versus domain knowledge in statistical forecasting of chaotic systems}, \href{https://doi.org/10.1103/physrevresearch.5.043252}{\emph{Phys. Rev. Res.} {\bfseries 5} (2023) 043252}.

\bibitem{NAN24}
S.~Nandy, B.~Mukherjee, A.~Bhattacharyya and A.~Banerjee, \emph{Quantum state complexity meets many-body scars}, \href{https://doi.org/10.1088/1361-648x/ad1a7b}{\emph{J. Phys. Condens. Matter} {\bfseries 36} (2024) 155601}.

\bibitem{PAL23}
K.~Pal, A.~Gill and T.~Sarkar, \emph{Time evolution of spread complexity and statistics of work done in quantum quenches}, \href{https://doi.org/10.1103/physrevb.108.104311}{\emph{Phys. Rev. B} {\bfseries 108} (2023) 104311}.

\bibitem{BHA22}
B.~Bhattacharjee, S.~Sur and P.~Nandy, \emph{Probing quantum scars and weak ergodicity breaking through quantum complexity}, \href{https://doi.org/10.1103/physrevb.106.205150}{\emph{Phys. Rev. B} {\bfseries 106} (2022) 205150}.

\bibitem{GAU23}
M.~Gautam, K.~Pal, K.~Pal, A.~Gill, N.~Jaiswal and T.~Sarkar, \emph{Spread complexity evolution in quenched interacting quantum systems}, {\emph{arXiv} (2023) }.

\bibitem{NIE06a}
M.A.~Nielsen, M.R.~Dowling, M.~Gu and A.C.~Doherty, \emph{Quantum {Computation} as {Geometry}}, \href{https://doi.org/10.1126/science.1121541}{\emph{Science} {\bfseries 311} (2006) 1133}.

\bibitem{BHA24}
A.~Bhattacharya, R.N.~Das, B.~Dey and J.~Erdmenger, \emph{Spread complexity for measurement-induced non-unitary dynamics and {Zeno} effect}, \href{https://doi.org/10.1007/jhep03(2024)179}{\emph{J. High Energy Phys.} {\bfseries 2024} (2024) 179}.

\bibitem{BAL23a}
V.~Balasubramanian, J.M.~Magan and Q.~Wu, \emph{Quantum chaos, integrability, and late times in the {Krylov} basis}, \href{https://doi.org/10.48550/arxiv.2312.03848}{\emph{arXiv} (2023) }.

\bibitem{LIN22}
H.W.~Lin, \emph{The bulk {Hilbert} space of double scaled {SYK}}, \href{https://doi.org/10.1007/jhep11(2022)060}{\emph{J. High Energy Phys.} {\bfseries 2022} (2022) 60}.

\bibitem{RAB23}
E.~Rabinovici, A.~S{\'{a}}nchez-Garrido, R.~Shir and J.~Sonner, \emph{A bulk manifestation of {Krylov} complexity}, \href{https://doi.org/10.1007/jhep08(2023)213}{\emph{J. High Energy Phys.} {\bfseries 2023} (2023) 213}.

\bibitem{HUH24}
K.~Huh, H.~Jeong and J.F.~Pedraza, \emph{Spread complexity in saddle-dominated scrambling}, \href{https://doi.org/10.1007/jhep05(2024)137}{\emph{J. High Energy Phys.} {\bfseries 2024} (2024) 137}.

\bibitem{NIE11}
M.A.~Nielsen and I.L.~Chuang, \emph{Quantum Computation and Quantum Information: 10th Anniversary Edition}, Cambridge University Press (2011).

\bibitem{SHO94}
P.W.~Shor, \emph{Algorithms for quantum computation: discrete logarithms and factoring}, IEEE (Nov., 1994), \href{https://doi.org/10.1109/sfcs.1994.365700}{10.1109/sfcs.1994.365700}.

\bibitem{CIN24}
S.~{\v C}indrak, B.~Donvil, K.~L{\"{u}}dge and L.~Jaurigue, \emph{Enhancing the performance of quantum reservoir computing and solving the time-complexity problem by artificial memory restriction}, \href{https://doi.org/10.1103/physrevresearch.6.013051}{\emph{Phys. Rev. Res.} {\bfseries 6} (2024) 013051}.

\bibitem{MUJ23}
P.~Mujal, R.~Mart{\'{\i}}nez-Pe{\~{n}}a, G.L.~Giorgi, M.C.~Soriano and R.~Zambrini, \emph{Time-series quantum reservoir computing with weak and projective measurements}, {\emph{npj Quantum Inf.} {\bfseries 9} (2023) 16}.

\bibitem{LI15}
Z.~Li, X.~Liu, N.~Xu and J.~Du, \emph{Experimental {Realization} of a {Quantum} {Support} {Vector} {Machine}}, \href{https://doi.org/10.1103/physrevlett.114.140504}{\emph{Phys. Rev. Lett.} {\bfseries 114} (2015) 140504}.

\bibitem{SAG21}
V.~Saggio, B.E.~Asenbeck, A.~Hamann, T.~Str{\"{o}}mberg, P.~Schiansky, V.~Dunjko et~al., \emph{Experimental quantum speed-up in reinforcement learning agents}, \href{https://doi.org/10.1038/s41586-021-03242-7}{\emph{Nature} {\bfseries 591} (2021) 229}.

\bibitem{DAM12}
J.~Dambre, D.~Verstraeten, B.~Schrauwen and S.~Massar, \emph{Information processing capacity of dynamical systems}, \href{https://doi.org/10.1038/srep00514}{\emph{Sci. Rep.} {\bfseries 2} (2012) 514}.

\bibitem{MAR23}
R.~Mart{\'{\i}}nez-Pe{\~{n}}a, J.~Nokkala, G.L.~Giorgi, R.~Zambrini and M.C.~Soriano, \emph{Information {Processing} {Capacity} of {Spin}-{Based} {Quantum} {Reservoir} {Computing} {Systems}}, \href{https://doi.org/10.1007/s12559-020-09772-y}{\emph{Cogn. Comput.} {\bfseries 15} (2023) 1440}.

\bibitem{SCH15a}
M.~Schuld, I.~Sinayskiy and F.~Petruccione, \emph{An introduction to quantum machine learning}, \href{https://doi.org/10.1080/00107514.2014.964942}{\emph{Contemp. Phys.} {\bfseries 56} (2015) 172}.

\bibitem{MAT21}
D.~Mattern, D.~Martyniuk, H.~Willems, F.~Bergmann and A.~Paschke, \emph{Variational {Quanvolutional} {Neural} {Networks} with enhanced image encoding}, \href{https://doi.org/10.48550/arxiv.2106.07327}{\emph{arXiv} (2021) }.

\bibitem{MCC16}
J.R.~McClean, J.~Romero, R.~Babbush and A.~Aspuru-Guzik, \emph{The theory of variational hybrid quantum-classical algorithms}, \href{https://doi.org/10.1088/1367-2630/18/2/023023}{\emph{New J. Phys.} {\bfseries 18} (2016) 023023}.

\bibitem{SCH21}
M.~Schuld, R.~Sweke and J.J.~Meyer, \emph{Effect of data encoding on the expressive power of variational quantum-machine-learning models}, \href{https://doi.org/10.1103/physreva.103.032430}{\emph{Phys. Rev. A} {\bfseries 103} (2021) 032430}.

\end{thebibliography}\endgroup
\newpage

\appendix
\section{Hamiltonians}
\label{sec:appendix}\

\begin{align}
    H_1 &=  0.5X_1 \nonumber \\
    H_2 &=  0.5(X_1 + X_2) \nonumber \\
    H_3 &=  0.5( X_1 + X_2+ X_3) \nonumber \\
    H_4 &=  0.5(X_1 + X_2 + X_3 + X_4) \nonumber \\
    H_{I1} &= 0.5\sum_{i=1,j>i}^{3}X_iX_j+0.5\sum_{i=0}^3Z_i \nonumber \\
    H_{I2} &= 0.4X_1X_2+0.5\left(\sum_{j=3}^4X_1X_j + \sum_{i=2,j>i}^{4}X_iX_j+\sum_{i=1}^4Z_i\right) \nonumber \\
    H_{I3} &= 0.35X_1X_2+0.4X_1X_3+0.45X_1X_4+0.6X_2X_3 \nonumber \\
    & \quad + 0.55X_2X_4 + 0.5X_3X_4 + 0.5 \sum_{i=1}^4Z_i \nonumber 
\end{align}

\end{document}